\documentclass[preprint,showpacs,preprintnumbers,amsmath,amssymb, superscriptaddress,longbibliography,nofootinbib]{revtex4-1}
\usepackage{graphicx} 
\usepackage{xcolor}
\usepackage{amsmath}
 \usepackage{hyperref}
\usepackage[top = 2cm, bottom = 2cm, right = 2cm, left =2cm]{geometry}

\begin{document}

\title{Localized five-dimensional rotating brane-world black hole Analytically Connected to an to an AdS$_5$ boundary.}

\author{Milko Estrada }
\email{milko.estrada@gmail.com}
\affiliation{Departamento de Física, Facultad de Ciencias, Universidad de Tarapacá, Casilla 7-D, Arica, Chile}

\author{Francisco Tello-Ortiz }
\email{francisco.tello@ufrontera.cl}
\affiliation{Departamento de Ciencias Físicas, Universidad de La Frontera, Casilla 54-D, 4811186 Temuco, Chile.}

\date{\today}

\begin{abstract}
We provide a method to describe the geometry of an analytic, exponentially localized $5D$ rotating braneworld black hole, using the $5D$ Janis–Newman algorithm in Hopf coordinates. The induced metric on the brane matches the standard $4D$ Kerr spacetime. Two curvature singularities arise: one confined to the $3$-brane at $z = r = 0$, and another that, on the brane, reproduces the Kerr singularity at $r = 0$, $\bar{\theta} = \pi/2$. The inner and event horizons, together with the stationary limit hypersurfaces, extend into the extra dimension in a pancake-like shape. We describe their behavior in the bulk.

The energy–momentum tensor represents a source transitioning from an anisotropic, non-diagonal structure to a vacuum with negative cosmological constant. Thus, the localized black hole connects to an AdS$_5$ boundary. The geometry is supported by a non-diagonal anisotropic fluid in the bulk, requiring no matter on the brane. To evaluate the energy conditions, we use a one-form from the dual basis that yields a diagonal energy–momentum tensor. The energy conditions are satisfied close to the brane, while they are violated at a location outside the brane but within the extension of the event horizon. The latter is required to support the rotating geometry.

\end{abstract}

\maketitle

\section{Introduction}

In recent years, there has been growing interest in investigating black holes within the context of extra dimensions. Several studies \cite{Cai:2020igv,LIGOScientific:2016kms,LIGOScientific:2017vwq} suggest that, at energy scales around 1 TeV, the formation of higher-dimensional black holes could be a realistic possibility in future particle colliders. Furthermore, the AdS/CFT correspondence \cite{Maldacena:1997re} has become a powerful framework for studying black holes in higher-dimensional spacetimes. In particular, \cite{Cai:2020igv} notes that a weakly coupled gravitational theory in a five-dimensional AdS spacetime is dual to a strongly coupled gauge theory in four dimensions.

The recent detection of gravitational waves produced by the collision of rotating black holes \cite{LIGOScientific:2016aoc} has also stimulated the theoretical study of these objects. In this context, rotating black holes in higher dimensions have been explored from a theoretical perspective. For instance, references \cite{Shaymatov:2018fmp,Shaymatov:2020tna} suggest that the solution for a rotating black hole in five dimensions exhibits an intriguing feature: it can be overspun under linear accretion. That is, it could be transformed into a naked singularity through overcharging or excessive rotation, thereby violating the Cosmic Censorship Conjecture. This occurs when the black hole possesses two angular momenta, but not when it has only one.

Within the framework of the study of gravitational scenarios in higher dimensions, braneworld models have attracted significant attention in recent decades. The pioneering Randall–Sundrum (RS) model \cite{Randall:1999vf,Randall:1999ee} offers a geometric explanation for why gravity is weaker than the other fundamental forces of nature. See appendix \ref{ResenaRS}.  Braneworld models have been theoretically tested using data from the LIGO/Virgo collaboration in recent years \cite{Banerjee:2021aln, Banerjee:2022jog, Visinelli:2017bny, Mishra:2021waw}. These models have also been confronted with LHC data from a theoretical standpoint. In this context, reference \cite{Casadio:2012pu} claims that a minimal mass for semiclassical microscopic black holes can be derived, which has a relevant impact on the description of black hole events at the LHC. Additionally, see reference \cite{daRocha:2006ei}, where the possible effects of the braneworld scenario on mini black holes and their measurable consequences in observations at the Large Hadron Collider are investigated. For more recent studies contrasting braneworld scenarios with the LHC, see references \cite{Das:2020gie, Belis:2023atb}.

It is worth noting that approaches to studying the representation of black holes (BHs) in a braneworld setup are primarily based on two methodologies: one originating from the brane and the other from the bulk. In the first methodology, four-dimensional black hole solutions are constructed by analyzing the equations of motion induced on the brane \cite{Shiromizu:1999wj}. Examples of rotating braneworld black holes obtained through this method can be found in Refs. \cite{Aliev:2005bi,Aliev:2009cg,Bohra:2023vls}. However, this approach does not fully capture the influence of the bulk geometry on the physical and geometric properties of the four-dimensional spacetime and, among other limitations, does not allow for a complete analysis of singularities in the extra dimension. 

The second approach to studying braneworld black hole models constructs spacetimes directly from the bulk, with the brane equations subsequently derived from it. This work focuses on this latter perspective. Usually, the representation of static black hole braneworld models takes the form of the line element \cite{Chamblin:1999by}
\begin{equation} \label{ElementodeLineaUsualBH}
    ds^2 = \frac{1}{(k |z| + 1)^2} \left( -f(r)dt^2 + \frac{dr^2}{f(r)} + r^2 d\Omega^2 + dz^2 \right)
\end{equation}

Thus, the metric tensor is factorizable, such that it corresponds to the multiplication of a warp factor by a five-dimensional spacetime. This five-dimensional spacetime typically corresponds to a static four-dimensional black hole embedded in an extra dimension, which is sometimes referred to as a ``black string". This form of geometric representation has also been studied for the case where the four dimensional black hole geometry is of a rotating nature \cite{Modgil:2001hm,Sengupta:2002fr,Biggs:2021iqw}.

In recent years, the study of braneworld black hole models from the perspective of the bulk equations of motion (rather than the equations induced on the brane) has regained interest, particularly since the emergence of the alternative methodology provided in Ref. \cite{Nakas:2020sey}. Unlike the setup described in the previous paragraph, this approach leads to a factorizable spacetime as well, but where the factor multiplying the warp factor now corresponds to a five-dimensional static black hole. As indicated in that reference, this allows for the representation of the geometry of an analytic, exponentially localized five-dimensional braneworld black hole. An interesting consequence of this representation is that the event horizon extends along the extra dimension, while the singularity remains localized on the brane. See some recent applications of this methodology in Refs. \cite{Nakas:2021srr,Neves:2021dqx,Nakas:2023yhj,Estrada:2024lhk,Neves:2024zwi,Pappas:2024qwm}. In this way, it becomes interesting to extend this methodology to represent rotating five-dimensional black holes.

Before proceeding, it is important to note the following: in four dimensions, rotating solutions are typically generated by applying the standard 4D Janis–Newman algorithm to static 4D black holes. However, the application of the algorithm in five dimensions is not trivial.  The reference \cite{Erbin:2014lwa} provided a five-dimensional version of this algorithm. Remarkably, it pointed out that by rewriting the angular components of the cross-section of a static five-dimensional black hole (BH), corresponding to an $S^3$ sphere, using Hopf bifurcation, it is possible to derive a five-dimensional version of the Janis–Newman algorithm. Consequently, by applying this algorithm to static five-dimensional solutions, rotating solutions can be obtained.

For all the reasons mentioned above, it is of physical interest, using the methodology previously described in Ref. \cite{Nakas:2020sey}, to represent the geometry of an analytic, exponentially localized five-dimensional rotating braneworld black hole. This, in contrast to the five-dimensional static braneworld black hole, which has been the focus of studies to date \cite{Nakas:2021srr,Neves:2021dqx,Nakas:2023yhj,Nakas:2020sey,Neves:2024zwi,Pappas:2024qwm}. In this work, we will address this problem by representing the geometry of an analytic, exponentially localized five-dimensional braneworld black hole. We will propose a methodology for this, using the five-dimensional version of the Janis-Newman algorithm in Hopf coordinates. We will study the behavior of the singularity and the asymptotic structure. We will test how the outer and event horizons, as well as the stationary limit hypersurfaces, extend along the extra dimension. Given that the energy-momentum tensor is non-diagonal and that the presence of the warp factor leads to highly non-linear components, we will use a one-form of the dual basis such that a diagonal energy-momentum tensor is obtained for a specific choice of one of the angular coordinates. In this way, we will test the energy conditions in the bulk.

\section{The Methodology Revisited}
In this section, we provide a brief review of the methodology recently developed in reference \cite{Nakas:2020sey} in order to represent a five-dimensional surface localized in a braneworld setup.

The Randall-Sundrum spacetime (See a brief overview of this model in Appendix \ref{ResenaRS}), which corresponds to a globally AdS space and has one (or two) branes embedded in the extra coordinate at $z = 0$, can be represented by the following line element:
\begin{equation} \label{ElementodeLineaZ}
    ds^2 = \frac{1}{(k |z| + 1)^2} \left( -dt^2 + dr^2 + r^2 d \bar{\Omega}^2 + dz^2 \right)
\end{equation}
where the brane coordinates are $(x^\mu=t,r,\bar{\theta},\bar{\phi})$ and where $d\bar{\Omega}^2 = d\bar{\theta}^2 + \sin^2 \bar{\theta} \, d \bar{\phi}^2 $. 

The following coordinate transformation was proposed in reference \cite{Nakas:2020sey}:

\begin{equation} \label{TransformacionesR}
r = \rho \sin \bar{\chi}, \quad z = \rho \cos \bar{\chi}
\end{equation}
where \( \bar{\chi} \in [0, \pi] \). The inverse transformation is:
\begin{equation} \label{RhoRZ}
    \rho = \sqrt{r^2 + z^2}, \quad \tan \bar{\chi} = \frac{r}{z}
\end{equation}
Thus, under this change of coordinates, the line element \eqref{ElementodeLineaZ} is:
\begin{equation} \label{ElementoDeLineaAngular}
    ds^2 = \frac{1}{(1 + k \rho |\cos \bar{\chi}|)^2} \left( -dt^2 + d\rho^2 + \rho^2 d \bar{\Omega}_3^2 \right)
\end{equation}
where
\begin{equation}
d\bar{\Omega}_3^2 = d\bar{\chi}^2 + \sin^2 \bar{\chi} \, d\bar{\theta}^2 + \sin^2 \bar{\chi} \sin^2 \bar{\theta} \, d\bar{\phi}^2
\end{equation}

The equation \eqref{ElementoDeLineaAngular} represents to a  localized five dimensional flat space-time in spherical coordinates, where the slice located at $\bar{\chi}=\pi/2$ and $z=0$ corresponds to a $3-$brane. 

\begin{figure}[h]   
    \centering 
    \includegraphics[scale=.8]{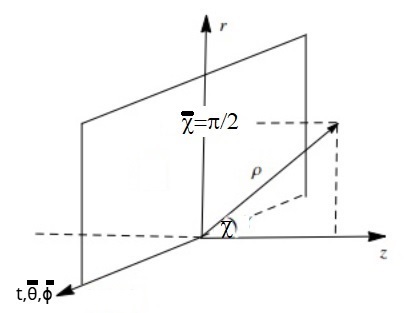} 
    \caption{brane world set up} \label{FigBrana}
 \end{figure}

Since $\sin \bar{\chi} \geq 0$ for $\chi \in [0, \pi]$, the coordinate $\rho \in [0, \infty]$. On the other hand, for $\bar{\chi} \in (\pi/2, \pi]$, where $\cos \bar{\chi} < 0$, it corresponds to the left side of the brane, while for $\bar{\chi} \in [0, \pi/2)$, where $\cos \bar{\chi} > 0$, it corresponds to the right side of the brane. Moreover, the spacetime preserves the $Z_2$ symmetry under rotations $\bar{\chi} \to \pi - \bar{\chi}$. See figure \ref{FigBrana}. 

Taking into account the symmetry of the line element under the transformation $\bar{\chi} \to \pi - \bar{\chi}$, it is sufficient to consider only one of the two possible $\bar{\chi}$ regimes. Therefore, Following reference \cite{Nakas:2020sey}, we will focus on the regime $\bar{\chi} \in [0, \pi/2]$, for which $\cos \bar{\chi} \geq 0$.

Thus, taking the above into account and in order to represent a five-dimensional localized black hole, reference \cite{Nakas:2020sey} replaces the two-dimensional flat part ($-dt^2 + d\rho^2$) of the line element in Eq. \eqref{ElementoDeLineaAngular} with the corresponding part of the four-dimensional Schwarzschild solution, such that the line element takes the following form:
\begin{align} \label{ElementoDeLineaSinValorAbsoluto}
    ds^2 &= \frac{1}{(1 + k \rho \cos \bar{\chi})^2} \left( -f(\rho) dt^2 + \frac{d\rho^2}{f(\rho)} + \rho^2 d\bar{\Omega}_3^2 \right) \\
         &= \text{WF} \cdot d\bar{s}_{S5D}^2
\end{align}
where $f(\rho) = 1 - \frac{2M}{\rho}$.

A remarkable feature of this methodology lies in the fact that the metric induced at the brane location at $\bar{\chi} = \pi/2$ and $z = 0$ corresponds to a four-dimensional Schwarzschild black hole. As a consequence of a five-dimensional black hole in the ambient metric structure, although the singularity is located at $\rho = 0$, i.e., at $r = z = 0$, the event horizon is not located at $z = 0$. Instead, it extends in the extra coordinate in a pancake-like shape. This suggests that thermodynamic quantities, such as entropy, might also extend in the extra coordinate.

\section{The new rotating case}

In this section, we provide a methodology to represent a five-dimensional braneworld model, corresponding to the product of a warp factor and a five-dimensional rotating black hole. To obtain the aforementioned rotating factor, we will apply the five-dimensional Janis–Newman algorithm \cite{Erbin:2014lwa} to the geometric factor previously described, corresponding to a five-dimensional static black hole.

It is important to note that in the previous section, the standard spherical coordinates are denoted by \{t, $\rho$, $\bar{\chi}$, $\bar{\theta}$, $\bar{\phi}$\}. In this section, we denote the Hopf coordinates as \{t, $\rho$, $\chi$, $\theta$, $\phi$\}. Below, we will describe the coordinate transformation relations and the domain of each coordinate.

As previously mentioned, since the resulting field equations are highly nonlinear and nontrivial when the cross-section is expressed in standard spherical coordinates, the $5D$ version of the JN algorithm relies on the angular cross-section of the seed static 5D solution being written in Hopf coordinates \cite{Erbin:2014lwa}. 

Using the fact that the relationship between spherical and Hopf coordinates is given by:
\begin{equation} \label{AngulosWF}
    \cos \bar{\chi}=\cos \theta \sin \chi
\end{equation}

It is direct to check that equation \eqref{TransformacionesR}  takes the following form in Hopf coordinates
\begin{align} 
    &z= \rho \cos \theta \sin \chi \label{TransformacionesHopf} \\
    &r^2=\rho^2 \left ( 1- \cos ^2\theta \sin ^2 \chi\right) \label{TransformacionesHopf1} \\
    &\rho^2=z^2+r^2 \label{TransformacionesHopf2}
\end{align}

In this way, the right-hand sides of equation \eqref{RhoRZ} now depend on the form taken by the coordinates $z, r$ in equations \eqref{TransformacionesHopf},\eqref{TransformacionesHopf1}.

The warp factor (WF) takes the form given in equation \eqref{ElementoDeLineaSinValorAbsolutoHopf}. We will begin by rewriting the line element in equation \eqref{ElementoDeLineaSinValorAbsoluto}, where the cross-section is now written in Hopf coordinates:

\begin{align} \label{ElementoDeLineaSinValorAbsolutoHopf}
    ds^2 &=  \frac{1}{(1 + k \rho \cos \theta \sin \chi)^2} \left( -f(\rho) dt^2 + \frac{d\rho^2}{f(\rho)} + \rho^2 d\Omega_3^2 \right) \\
         &= \text{WF} \cdot ds_{S5D}^2
\end{align}
where
\begin{equation}
  d\Omega_3^2=   d\chi^2 + \sin^2 \chi \, d\theta^2 \nonumber + \cos^2 \chi \, d\phi^2
\end{equation}
where $0 \leq \chi \leq \frac{\pi}{2}$ and $0 \leq \theta, \phi \leq 2\pi$. We can see a review of Hopf coordinates in references \cite{Sakaguchi:2005mz,Erbin:2014lwa}. 

It is worth noting that, as mentioned earlier, we are considering the right-hand side of the spacetime such that $\cos \bar{\chi} \geq 0$, i.e. $\bar{\chi} \in [0,\pi/2]$. Taking into account that the domain in Hopf coordinates is given by $\chi \in [0,\pi/2]$, we will consider values of the variable $\theta \in [0,\pi/2]$. In this way, the warp factor satisfies $\mbox{WF} \sim \sin \chi \cos \theta \geq 0$. As mentioned earlier, it is very difficult to draw a schematic representation in Hopf coordinates.

Following the five-dimensional Janis–Newman algorithm of reference \cite{Erbin:2014lwa}, and using the following ansatz :
\begin{equation}
    f(\rho) = 1-m(\rho)/\rho^2
\end{equation}

As we will see below, in order for the induced metric to correspond to a four-dimensional Kerr spacetime, in this work we will study the case where:
\begin{equation} \label{FuncionMasaMia}
    m(\rho)=M \rho
\end{equation}
where the parameter $M$ is related to the mass of the rotating black hole.

The function $m(\rho)$ is sometimes referred to as the mass function. It is worth emphasizing that the form of our mass function, given in Eq.\eqref{FuncionMasaMia}, differs from that used in the static Schwarzschild–Tangherlini solution in 5D, where $m(\rho) = \text{constant}$ (which, when the 5D Janis–Newman algorithm is applied, leads to the rotating Kerr–Myers–Perry spacetime in 5D). In this regard, it should be noted that within our braneworld framework it is not possible to employ a constant mass function as in the aforementioned 5D Schwarzschild–Tangherlini case (or consequently in the 5D rotating Kerr–Myers–Perry case), since doing so would make it impossible to define the ADM mass in the induced 4D metric. Therefore, we observe that the structure of the static geometry of Eq. \eqref{ElementoDeLineaSinValorAbsoluto} and the rotating geometry of Eq. \eqref{ParteRotante} differ from those of the 5D Schwarzschild–Tangherlini and Kerr–Myers–Perry geometries, respectively.

We apply this algorithm to the right part of the factorizable line element \eqref{ElementoDeLineaSinValorAbsolutoHopf}, wich represents a $5D$ static black hole, $d{s}_{S5D}^2$, where $f(\rho)$ is now given by the last equation. Thus, we obtain the following line element:
\begin{equation} \label{ElementoDeLineaAmbiente}
    ds^2= \frac{1}{(1 + k \rho \cos \theta \sin \chi)^2} \cdot d{s}_{R5D}^2
\end{equation}
where from 
\begin{align} \label{ParteRotante}
    d{s}_{R5D}^2 =& -dt^2 + \frac{m(\rho)}{\Sigma^2} \left( dt - \omega \right)^2 + \rho^2 \frac{\Sigma^2}{\Delta}  d\rho^2 + \Sigma^2 d\chi^2 + (\rho^2 + a^2)\sin^2 \chi d\theta^2 + (\rho^2 + b^2)\cos^2 \chi d \phi^2
\end{align}
\begin{align}
    \Sigma^2 =& \rho^2 + a^2 \cos^2 \chi + b^2 \sin ^2 \chi \\
    \Delta =& (\rho^2+a^2)(\rho^2+b^2)- m(\rho) \rho^2 \\
    \omega= &a \sin^2 \chi d \theta + b \cos ^2 \chi d \phi
\end{align}

The metric \eqref{ParteRotante} has three Killing vector fields,
$\left\{\frac{\partial}{\partial t}, \frac{\partial}{\partial \theta}, \frac{\partial}{\partial \phi}\right\}$, where the parameter $a,b$ are related to the conserved quantity associated with $\frac{\partial}{\partial \theta}$ $\frac{\partial}{\partial \phi}$, respectively
\cite{Sakaguchi:2005mz}. In the following of this work, for simplicity, we assume that the conserved charge associated with $\frac{\partial}{\partial \theta}$ vanishes, i.e., $a = 0$.

\subsection{The induced metric on the brane}

For convenience, we will rewrite the line element \eqref{ParteRotante} as follows:

\begin{align} \label{ParteRotante1}
    d{s}_{R5D}^2 =&- \left (1- \frac{m(\rho)}{\Sigma^2} \right ) dt^2 + \rho^2 \frac{\Sigma^2}{\Delta}  d\rho^2 - 2 \frac{m(\rho)}{\Sigma^2} b \cos^2 \chi dt d\phi + \frac{m(\rho)}{\Sigma^2} b^2 \cos^4 \chi d\phi^2 \nonumber \\
    &+ b^2 \sin^2 \chi d\chi^2+b^2 \cos^2 \chi d\phi^2 \nonumber \\
    &+ \rho^2 \left ( d\chi^2 + \sin^2 \chi \, d\theta^2 + \cos^2 \chi \, d\phi^2 \right)
\end{align}

In order to compare the previous expression written in Hofp coordinates with the usual spherical coordinates, we will use the following identities:
\begin{align}
    & d\chi^2 + \sin^2 \chi \, d\theta^2 + \cos^2 \chi \, d\phi^2 = d\bar{\chi}^2 + \sin^2 \bar{\chi} \, d\bar{\theta}^2 + \sin^2 \bar{\chi} \sin^2 \bar{\theta} \, d\bar{\phi}^2 \label{Identidad1} \\
    & \sin^2 \chi = \cos^2 \bar{\chi}+ \sin^2 \bar{\chi} \cos^2 \bar{\theta} \label{Identidad2}\\
    & \cos^2 \chi = \sin^2 \bar{\chi} \sin^2 \bar{\theta} \label{Identidad3} \\
    &\phi= \bar{\phi} \label{Identidad4}
\end{align}

At the brane location $\bar{\chi}=\pi/2$ we can note that:
\begin{align}
     & d\chi^2 + \sin^2 \chi \, d\theta^2 + \cos^2 \chi \, d\phi^2  =  d\bar{\theta}^2 + \sin^2 \bar{\theta} \, d\bar{\phi}^2 \\
     & \sin \chi = \cos \bar{\theta} \label{SenoChi} \\
     & \cos \chi= \sin \bar{\theta} \label{CosenoChi}  \\
     &\Rightarrow d\chi^2=d\bar{\theta}^2
\end{align}
Thus, at $\bar{\chi}=\pi/2$ and $z=0$:
\begin{align}
\Sigma^2 =& r^2 +  b^2 \cos ^2 \bar{\theta} =\bar{\Sigma}_{Kerr-4D} \\
    \Delta =& r^2 ( r^2+b^2 - M r ) = r^2 \bar{\Delta}_{Kerr-4D}  \\
    \omega= & b \sin ^2 \bar{\theta} d \bar{\phi}    
\end{align}

Taking into account that the warp factor of equation \eqref{ElementoDeLineaAmbiente} at the brane location $z=0$ is equal to unity, the induced metric on the brane is:
\begin{align}
    d\bar{s}^2 \big |_{brane}=&- \left( 1- \frac{Mr}{\bar{\Sigma}_{Kerr-4D}} \right) dt^2 + \frac{\bar{\Sigma}_{Kerr-4D}}{\bar{\Delta}_{Kerr-4D} } dr^2 + \left (  r^2+b^2 + \frac{Mrb^2}{\bar{\Sigma}_{Kerr-4D}} \sin^2 \bar{\theta} \right) d\bar{\phi}^2 \nonumber \\
   & + \bar{\Sigma}_{Kerr-4D} d\bar{\theta}^2 - \frac{2Mrb}{\bar{\Sigma}_{Kerr-4D}} \sin^2\bar{\theta} d\bar{\phi}dt
\end{align}

This precisely corresponds to a four-dimensional rotating Kerr spacetime. 

\subsection{The energy momentum tensor on the brane}
Following reference \cite{Nakas:2020sey}, the energy-momentum tensor is decomposed as follows:
\begin{equation}
    T_{MN} = T^{(B)}_{MN} + \delta^{\mu}_M \delta^{\nu}_N T^{(br)}_{\mu\nu} \delta(y)
\end{equation}
where $T^{(B)}_{MN}$ and $T^{(br)}_{\mu\nu}$ represent the energy-momentum tensor in the bulk and on the brane localization, respectively. The latter can be written as:
\begin{equation} \label{EnergiaMomentumBrana1}
    T^{(br)}_{\mu\nu} = -\sigma h_{\mu\nu} + \tau_{\mu \nu}
\end{equation}
where $\sigma$ represents the brane tension and where $\tau_{\mu \nu}$ represents the other possible sources on the brane. In reference \cite{Aliev:2005bi}, it was shown that in the rotating case, there are no matter fields on the brane, i.e. $\tau_{\mu \nu}=0$. Thus, the Israel conditions  lead to the energy-momentum tensor on the brane having the following form:
\begin{equation} \label{EnergiaMomentumBrana2}
    T^{(br)}_{\mu\nu} = -\frac{6k}{\kappa_5^2}h_{\mu \nu}
\end{equation}

It is straightforward to check that by comparing equations \eqref{EnergiaMomentumBrana1} and \eqref{EnergiaMomentumBrana2}, the brane tension is given by $\sigma = \dfrac{6k}{\kappa_5^2}$ with $\tau_{\mu \nu} = 0$. Thus, we can note that the five-dimensional geometry of a rotating braneworld black hole connected to an asymptotically AdS$_5$ space is supported by the fields of the energy momentum tensor present in the bulk, and it is not necessary to add additional sources for the energy on the brane and the four-dimensional geometry. 

\paragraph*{ About the effective field equations on the brane : }

Note that in reference \cite{Aliev:2005bi} it has already been shown that the field equations induced on the brane are such that the Einstein tensor vanishes, $G^\mu_\nu = 0$, with $(\mu,\nu=t,r,\bar{\theta},\bar{\phi})$. See the same in references \cite{Aliev:2009cg,Bohra:2023vls}. This is indeed the expected result since the induced line-element on the brane is described by the Kerr solution, which is a vacuum solution.

\subsection{Singularity of the $5D$ spacetime and asymptotic behavior}

The line element is:

\begin{equation} \label{ElementoDeLineaAmbiente1}
    ds^2= \frac{1}{(1 + k \rho \cos \theta \sin \chi)^2} \cdot d{s}_{R5D}^2
\end{equation}
where $d{s}_{R5D}^2$ is given by equation \eqref{ParteRotante} wit $a=0$. Thus, the Ricci scalar is:

\begin{align} \label{RicciAmbiente}
    &R= -20 k^2 +\frac{M}{\rho \left( b^2 \left ( \cos(2 \chi) - 1 \right ) - 2 \rho^2 \right)}\bigg (  3 \cos(2 \theta) \cos(2 \chi) k^2 \rho^2 - 3 \cos(2 \theta) k^2 \rho^2 + 3 \cos(2 \chi) k^2 \rho^2 - 3 k^2 \rho^2  \nonumber \\
    & + 12 \sin(-\theta + \chi) k \rho + 12 \sin(\theta + \chi) k \rho - 4 \bigg)
\end{align}

In the equation above, we can notice that there is a contribution from the bulk cosmological constant, $-20k^2$, plus a contribution proportional to the mass located on the brane, $M$. Regarding this, we can identify the following characteristics.

\begin{itemize}
    \item We can note that the first factor in the denominator of the second term diverges at $\rho = 0$.
    \item  we observe that the second factor in the denominator of the second term diverges at $\rho = 0$ and $\cos(2\chi)=1 \Rightarrow \chi=0$
\end{itemize}

However, since the first factor already diverges at $\rho = 0$, regardless of the value of $\chi$, this latter divergence is mandatory. Thus, there is a mandatory curvature singularity at $r=0,z=0$. Therefore, similarly to the static case \cite{Nakas:2020sey}, the black hole singularity is entirely confined to the 3-brane.

Regarding the divergence of the second factor in the denominator of the second term in equation \eqref{RicciAmbiente}, that is, at $\rho = 0$ and $\chi = 0$, we can point out the following:

\begin{itemize}
    \item At the brane location $\bar{\chi} = \pi / 2$, we see that equation \eqref{AngulosWF} is satisfied.
    \item  On the other hand, from equations \eqref{SenoChi} and \eqref{CosenoChi}, for this same brane location, we note that $cos(2\chi) = \sin^2(\bar{\theta}) - \cos^2(\bar{\theta})$, that is, $\bar{\theta} = \pi / 2$.
    \item This means that the divergence of this last factor at the brane location, $\bar{\chi} = \pi / 2$, is analogous to the divergence of the 4D Kerr solution at $r = 0$ and $\bar{\theta} = \pi / 2$. However, as mentioned above, the divergence discussed in the previous paragraph at $\rho = 0$ is unavoidable in our embedding.
\end{itemize}

On the other hand, we can test that the Ricci scalar tends to $-20k^2$ at infinity of the coordinate $\rho \to \infty$. This value matches the corresponding one for an AdS$_5$ space, representing $k$ as the inverse of the AdS radius. Thus, we can interpret our geometry as an exponentially localized five-dimensional brane-world rotating black hole connected to an AdS$_5$ boundary. Therefore, the AdS structure emerges either when we move far away from the brane, i.e., $z \to \infty$, or at large distances along the brane, i.e., $r \to \infty$.

On the other hand, from the apendix \ref{EMComponentes}, we can check that the asymptotic values of the Einstein tensor and the energy-momentum tensor are:
\begin{equation}
    \displaystyle \lim_{\rho \to \infty} G^A_B=\displaystyle \lim_{\rho \to \infty} T^A_B = 6 k^2 = - \Lambda_{5DAdS} \cdot \delta^A_B
\end{equation}
with $(A,B=t,\rho,\chi,\theta,\phi)$. Thus, we can note that the off-diagonal components of the energy-momentum tensor, $T_t^\phi$ and $T_\phi^t$, vanish asymptotically, while the asymptotic values of the diagonal components are equal to the value of a negative cosmological constant corresponding to a 5D Anti-de Sitter space. In this way, the energy-momentum tensor can be interpreted as a source of matter that transitions in the $\rho$ coordinate from a non-diagonal and anisotropic structure to a vacuum distribution characterized by the 5D cosmological constant. Thus, the ambient spacetime transitions from a 5D rotating black hole to an AdS space. Therefore, the asymptote can be interpreted as a geometric boundary, such that the geometry can be thought of as an analytic, exponentially localized five-dimensional brane-world rotating black hole connected to an AdS$_5$ boundary.

\subsection{Event Horizon} \label{SeccionHorizontes}

To compute the singularities of coordinates, where the event horizons are located, we use the condition $g^{rr} = 0 \Rightarrow \Delta = 0$.
\begin{align}
     \Delta =& \rho^2 \left ( \rho^2- M\rho +b^2 \right) =0
\end{align}
As mentioned before, the value $\rho = 0$ is associated with a curvature singularity. In this subsection, we are interested in event horizons associated with coordinate singularities. Thus, we solve the following equation:
\begin{align}
    &  \rho_\pm^2- M\rho_\pm +b^2  =0 \label{EcuacionParaHorizontes} \Rightarrow \\
    & \rho_\pm= \frac{1}{2} \left ( M \pm \sqrt{M^2-4b^2} \right) \,\,\,\, \mbox{, with $M>2b$} \label{RhoMasMenos}
\end{align}
In the case $M = 2b$, we are dealing with the extremal case. For $M > 2b$, we will analyze both branches. 

From equations \eqref{ElementoDeLineaAmbiente} and \eqref{ParteRotante1}, it is easy to check that $g_{\rho \rho}(\rho_{\pm}) \to \infty$; thus, the points $\rho_{\pm}$ correspond to the locations of the outer and inner horizons. However, we are also interested in determining the locations of the horizons in terms of the radial coordinate $r$. In this connection, using equations \eqref{RhoMasMenos}, \eqref{RhoRZ}, and \eqref{CambioCoordenadas}, we find that

\begin{equation}
    r_{\pm} = \sqrt{\frac{1}{4} \left ( M \pm \sqrt{M^2-4b^2}   \right)^2 - \frac{ \left( \exp (k |y|)-1 \right)^2}
    {k^2}}
\end{equation}

We note that $r_{\pm}$ depends on the parameters $M$, $k$, and $y_0$. On the 3-brane, where $y_0 = 0$, the values of $r_{\pm}$ are similar to those of the Kerr geometry, and this remains true independently of the value of $k$.

As we can see from the last equation, the inner and outer horizons extend along the extra coordinate $y$. From this equation, we can also notice that, as we move along the extra dimension, both horizons shrink exponentially fast and become zero at a certain distance:

\begin{equation}
|y_{0}^{(\pm)}| = \frac{1}{k} \ln \left(   1 + \frac{k}{2} \left ( M \pm \sqrt{M^2-4b^2} \right ) \right )
\end{equation}

In this way, the inner and outer horizons of the rotating black hole take the shape of a ``pancake", with their longer side extended along the brane and their shorter side stretching into the bulk (the extra-dimensional volume) over an exponentially small distance. 

\begin{figure}[ht]
  \begin{center}
      \includegraphics[width=4.5in]{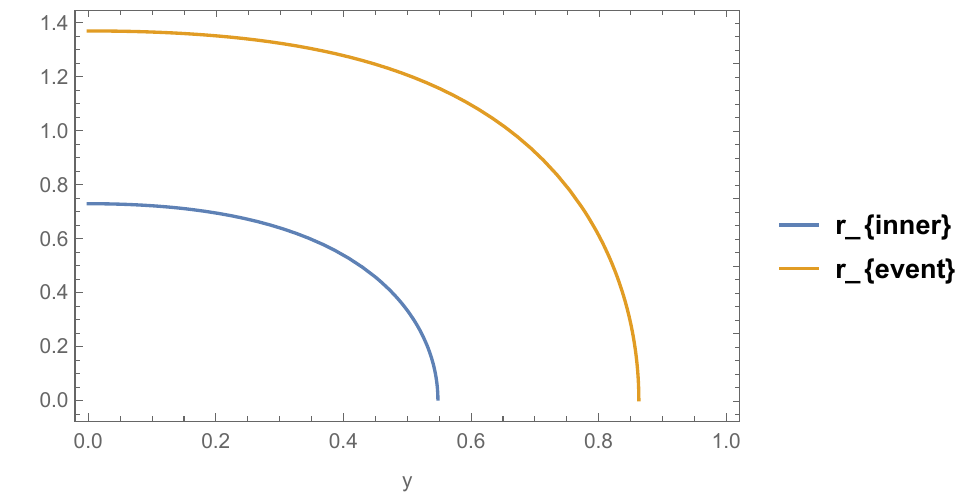}
        \caption{Behavior of the horizons $r_{\pm}$ for $k=b=1,M=2.1$}.\label{FigHorizontes}
  \end{center}
\end{figure}

On the other hand, we can numerically verify in Figure \ref{FigHorizontes} that the inner horizon vanishes more rapidly than the event horizon. This is a nontrivial result, since in recent years it has been discussed that the presence of an inner horizon may be associated with instability issues. Therefore, once the inner horizon disappears, a segment remains within the domain of the $y$ coordinate, where only an event horizon exists (without an inner one).

Analyzing another important property of a spacetime that describes the exterior of a rotating source, we focus on the existence of stationary limit hypersurfaces, for which $g_{tt} = 0$. These hypersurfaces can be found using equations \eqref{ElementoDeLineaAmbiente} and \eqref{ParteRotante1}

\begin{align}
    & \Sigma^2-M \rho=0 \nonumber \\
    &\left( \rho^{(s)}_{\pm} \right)^2-M \rho^{(s)}_{\pm} + b^2 \sin ^2 \chi =0 \label{CuadraticaErgosfera}
\end{align}

We can observe that Equation \eqref{CuadraticaErgosfera} is obtained by rearranging Equation \eqref{EcuacionParaHorizontes} such that $b^2 \to b^2 \sin^2 \chi$. Thus, using this rearrangement, we can infer that:

\begin{equation}
     \rho^{(s)}_{\pm} = \frac{1}{2} \left ( M \pm \sqrt{M^2-4b^2 \sin ^2 \chi} \right) \label{ErgosferaMasMenos}
\end{equation}

and consequently:
\begin{equation}
    r_{\pm}^{(s)} = \sqrt{\frac{1}{4} \left ( M \pm \sqrt{M^2-4b^2 \sin^2 \chi}   \right)^2 - \frac{ \left( \exp (k |y|)-1 \right)^2}
    {k^2}}
\end{equation}
where $r_{-}^{(s)}$ associated with the infinite redshift surface $S^-$ and $r^{(s)}_{+}$ is associated with the stationary limit surface (infinite redshift surface) $S^+$. On the other hand, we can notice that $r_{\pm}^{(s)}$ vanishes at:
\begin{equation} \label{DesvanecimientoSuperficie}
|(y_{0})_{\pm}^{(s)}| = \frac{1}{k} \ln \left(   1 + \frac{k}{2} \left ( M \pm \sqrt{M^2-4b^2 \sin^2 \chi} \right ) \right )
\end{equation}

\begin{figure}[ht]
  \begin{center}
      \includegraphics[width=5in]{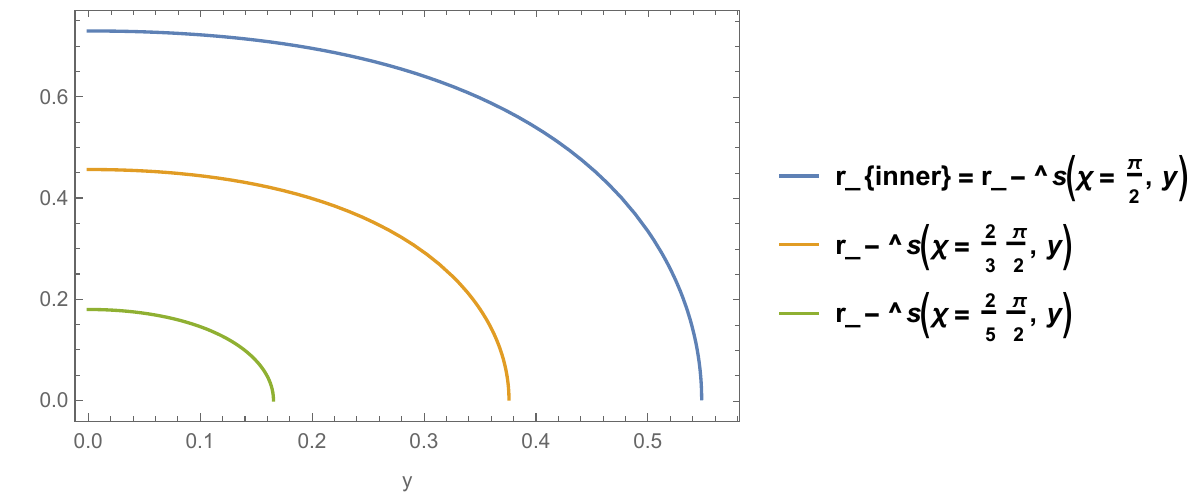}
      \includegraphics[width=5in]{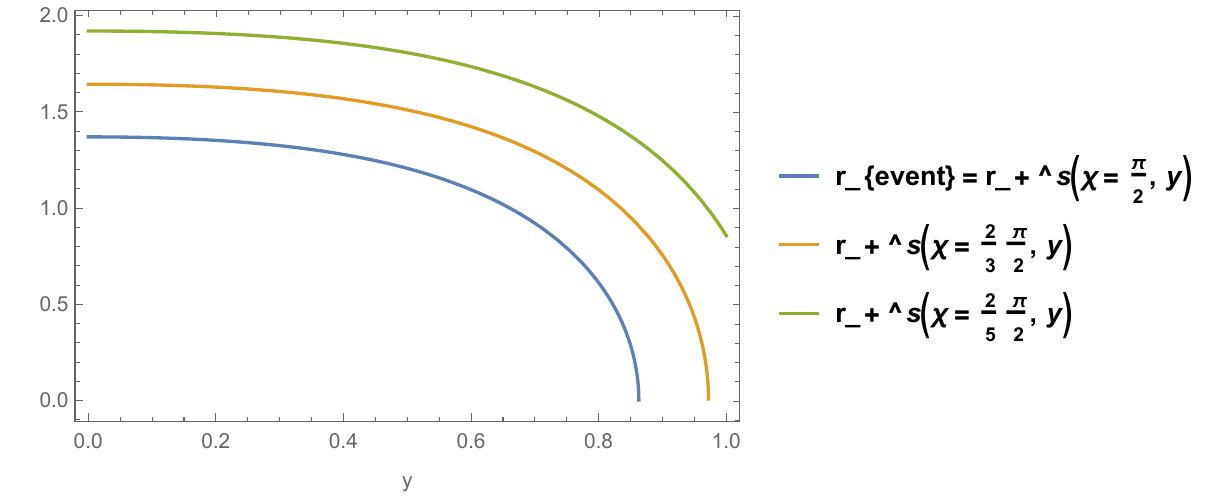}
        \caption{Behavior of the horizons $r_{\pm}$ vs $r^{(s)}_{\pm}$ for $k=b=1,M=2.1$}.\label{FigHorizontesSuperfices}
  \end{center}
\end{figure}

It is worth mentioning that, since we are using the Hopf coordinate representation, it becomes very difficult to draw a schematic representation of the geometry. Nevertheless, based on the above, we can outline the following:
\begin{enumerate}
    \item \label{Uno} The surface $r^{(s)}_{-}$ is entirely contained within the inner event horizon $r_{-}$ along the extra coordinate $y$. 
    \item \label{Dos} For $\sin \chi = 1$, $r_-$ and $r^{(s)}_-$ coincide, which indicates that these two hypersurfaces converge at this point.
\item  \label{Tres} We can check the last two previous points in the first panel of Figure \ref{FigHorizontesSuperfices}. In this panel, we can also notice that $r^{(s)}_{-}$ vanishes along the extra coordinate before the inner horizon does. As was mentioned, the analytical expression where the hypersurfaces vanish is given by Equation \eqref{DesvanecimientoSuperficie}.
    \item \label{Cuatro} $r_{+}$ is entirely contained within $r^{(s)}_{+}$ along the extra coordinate $y$.
    \item \label{Cinco} For $\sin \chi = 1$, $r_+$ and $r^{(s)}_+$ coincide, which indicates that these two hypersurfaces converge at this point.
\item \label{Seis} We can check the last two previous points in the second panel of Figure~\ref{FigHorizontesSuperfices}. In this panel, we can also notice that $r^{(s)}_{-}$ vanishes along the extra coordinate after the event horizon does.
\end{enumerate}

Thus, the spacetime studied in this work describes a five-dimensional rotating black hole solution, in which the central singularity is localized on our brane-universe, while both the horizons and the stationary limit hypersurfaces extend into the extra dimension.

On the other hand, it is worth mentioning that, at the location of the brane $\bar{\chi} = \pi/2$ and $z = 0$, using Equation \eqref{SenoChi} and \eqref{RhoRZ}, we arrive at 
\begin{equation}
     r^{(s)}_{\pm} = \frac{1}{2} \left ( M \pm \sqrt{M^2-4b^2 \cos ^2 \bar{\theta}} \right) \label{ErgosferaMasMenos}
\end{equation}
which, as expected, corresponds to the usual equation for determining the hypersurfaces in the 4D Kerr case.

\subsection{Energy conditions} \label{SeccionCondicionesEnergia}

The Null Energy Condition (NEC) requires that $D - p_i \geq 0$. The Weak Energy Condition (WEC), on the other hand, requires both $D - p_i \geq 0$ and $D \geq 0$. Therefore, in any region where the Weak Energy Condition (WEC) is satisfied, the Null Energy Condition (NEC) is automatically satisfied as well.

It is worth pointing out that, in our case, the rotating geometry is supported by a non-diagonal energy–momentum tensor, as shown in Appendix \ref{EMComponentes}. In this way, our matter sources differ from those in the static cases \cite{Nakas:2020sey,Nakas:2021srr,Neves:2021dqx}, where the energy–momentum tensor is diagonal.

The study of rotating spacetimes in 5D is non-trivial due to the fact that employing Hopf coordinates leads to a non-diagonal and highly non linear structure in both the equations of motion and the energy-momentum tensor. In the absence of the warp factor, the problem has been addressed by defining a one--form of the dual basis, such that a diagonal energy-momentum tensor is obtained for a specific location of one of the angular coordinates \cite{Amir:2020fpa,Estrada:2025sku}. In this work, we will use the basis from reference \cite{Estrada:2025sku} to obtain a diagonalized version of the energy-momentum tensor. As we will see below, the presence of the warp factor modifies the structure of the latter. This will allow us to analyze the energy conditions. This basis is given by:

\begin{align}
    e^{(t)} &= \sqrt{\bigg |g_{tt} - \frac{g_{t\phi}^2}{g_{\phi\phi}} - \Omega^2 \frac{g_{\theta\theta}g_{\phi\phi}-g_{\theta\phi}^2}{g_{\phi\phi} }\bigg |}dt=\sqrt{\pm \bigg (g_{tt} - \frac{g_{t\phi}^2}{g_{\phi\phi}} - \Omega^2 \frac{g_{\theta\theta}g_{\phi\phi}-g_{\theta\phi}^2}{g_{\phi\phi} }\bigg )}dt,\nonumber\\
    e^{(\rho)} &= \sqrt{g_{\rho\rho}}d\rho, \nonumber \\
    e^{(\chi)} &=\sqrt{g_{\chi\chi}}d\chi, \nonumber \\
    e^{(\theta)} &= - \sqrt{\frac{g_{\theta\theta}g_{\phi\phi}-g_{\theta\phi}^2}{g_{\phi\phi}}}(\Omega dt - d\theta),\nonumber \\
    e^{(\phi)} &= \sqrt{g_{\phi\phi}}\left( \frac{g_{t\phi}}{g_{\phi\phi}}dt + \frac{g_{\theta\phi}}{g_{\phi\phi}}d\theta + d\phi\right),
\end{align}
where: 

\begin{equation}
    \Omega = \frac{g_{t \phi}g_{\theta\phi}-g_{t\theta}g_{\phi \phi}}{g_{\theta\theta}g_{\phi\phi} - (g_{\phi\phi})^2}.
\end{equation}

The sign of the component $e^{(t)}$, either $+$ or $-$, is determined by whether the expression inside the parentheses is positive or negative. For an observer located beyond the event horizon, the appropriate choice is the negative sign. This is clearly illustrated in the static case, where $g_{tt} < 0$ in that region.

As a result, the effective energy-momentum tensor $T^{\mu\nu}$ can be interpreted as a projection determined by the selected tetrad basis, taking the form
\begin{equation}
    T^{(a)(b)} = e_{\mu}^{(a)} e_{\nu}^{(b)} T^{\mu\nu}.
\end{equation}
We then exploit the relationship between $T^{\mu\nu}$ and the Einstein tensor to express the energy-momentum tensor in the tetrad frame, $T^{(a)(b)}$, as
\begin{equation} \label{TensorEMDiagonalizado}
    8\pi T^{(a)(b)} = e_{\mu}^{(a)} e_{\nu}^{(b)} G^{\mu\nu}.
\end{equation}
Thus, by setting $\chi = \pi/2$, we note from Equation \eqref{AngulosWF} that for $\chi = \pi/2 \Rightarrow \cos \bar{\chi} = \cos \theta \Rightarrow \bar{\chi} = \theta$. We also use equations \eqref{TransformacionesHopf}, \eqref{TransformacionesHopf1}, \eqref{TransformacionesHopf2}, and \eqref{TensorEMDiagonalizado}. In this way, the components of the diagonalized tensor are written in a more compact form, as functions of $(r,z)$, that is, $T^{(a)(b)} = T^{(a)(b)}(r,z)$. Thus, the non-vanishing components are

\begin{align}
D=    8\pi T^{(t)(t)}=&\frac{
-12 b^{4} k^{2} \sqrt{r^{2}+z^{2}}
+ b^{2} \left(M - 7 k M z + 4 k^{2} M z^{2} - 24 k^{2} \left(r^{2}+z^{2}\right)^{3/2}\right)
}{
2 \sqrt{r^{2}+z^{2}} \, \left(b^{2}+r^{2}+z^{2}\right)^{2}
} \nonumber \\
&- \frac{3 \left(r^{2}+z^{2}\right)\left(M(-1+3kz)+4k^{2}\left(r^{2}+z^{2}\right)^{3/2}\right)}{
2 \sqrt{r^{2}+z^{2}} \, \left(b^{2}+r^{2}+z^{2}\right)^{2}
}
\end{align}

\begin{align}
p_\rho=  8\pi T^{(\rho)(\rho)}=&\frac{12b^4k^2\sqrt{r^2 + z^2} + 3(r^2 + z^2)\left( M(-1 + 3kz) + 4k^2(r^2 + z^2)^{3/2} \right) }{2\sqrt{r^2 + z^2}(b^2 + r^2 + z^2)^2} \nonumber \\
  & + \frac{b^2\left(24k^2(r^2 + z^2)^{3/2} + M(-1 + 7kz - 4k^2z^2) \right)}{2\sqrt{r^2 + z^2}(b^2 + r^2 + z^2)^2}
\end{align}

\begin{align}
 p_\chi=   8 \pi T^{(\chi)(\chi)}=&\frac{6b^4k^2\sqrt{r^2 + z^2} - b^2(M - 4kMz + k^2Mz^2 - 12k^2(r^2 + z^2)^{3/2})}{\sqrt{r^2 + z^2}(b^2 + r^2 + z^2)^2} \nonumber \\
                          &+ \frac{3k(r^2 + z^2)(2k(r^2 + z^2)^{3/2} + M(z - kz^2))}{\sqrt{r^2 + z^2}(b^2 + r^2 + z^2)^2}
\end{align}

\begin{align}
 p_\theta=   8 \pi T^{(\theta) (\theta)}=3k \left( 2k + \frac{Mz(1-kz)}{\sqrt{r^{2}+z^{2}}(b^{2}+r^{2}+z^{2})} \right)
\end{align}

\begin{align}
p_\phi=  T^{(\phi) (\phi)} =& \frac{6 b^{4} k^{2} \sqrt{r^{2}+z^{2}} 
- b^{2}\left(M - 4kMz + k^{2}Mz^{2} - 12k^{2}(r^{2}+z^{2})^{3/2}\right) 
 }
{\sqrt{r^{2}+z^{2}}\,(b^{2}+r^{2}+z^{2})^{2}} \nonumber \\
                   &+ \frac{ 3k(r^{2}+z^{2})\left(2k(r^{2}+z^{2})^{3/2} + M(z-kz^{2})\right)}{\sqrt{r^{2}+z^{2}}\,(b^{2}+r^{2}+z^{2})^{2}}
\end{align}

From the above, we can observe that:

\begin{equation} \label{AsintotaCcosmologica}
    \displaystyle \lim_{z \to \pm \infty} \left ( -D=p_\rho=p_\chi=p_\theta=p_\phi  \right)=6k^2 \sim -\Lambda
\end{equation}

This behavior maintains the AdS$_5$ gravitational background far from the brane, consistent with the asymptotic properties of AdS geometry. Moreover, it asymptotically gives rise to the exponential warp factor $e^{-k|y|}$ in the bulk spacetime, thereby reproducing the Randall--Sundrum model.

\begin{figure}[ht]
  \begin{center}
      \includegraphics[width=5in]{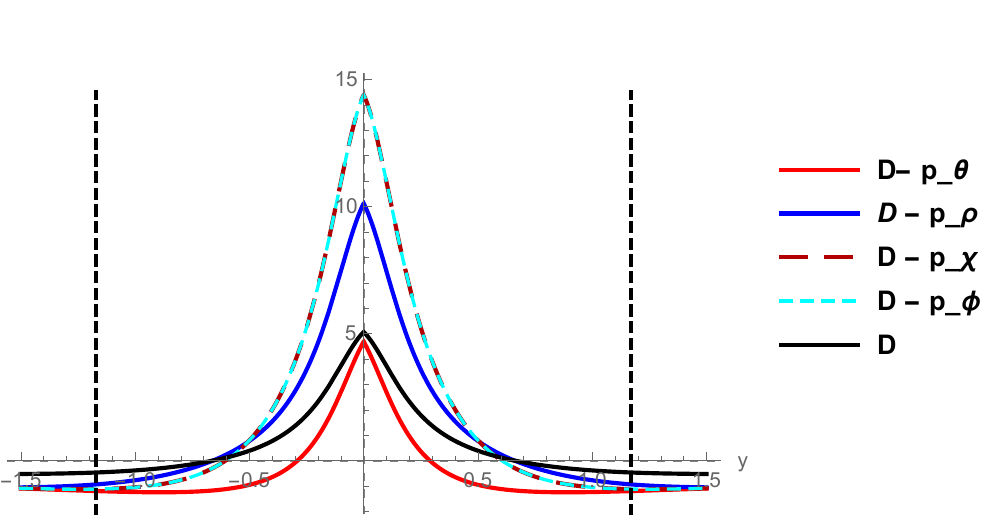}
         \caption{Energy conditions for $b=1,k=0.25, M=2.1,r=0.2$}.\label{FigCondicionesDeEnergia}
  \end{center}
\end{figure}

Following the discussion in the previous paragraphs and using equation \eqref{CambioCoordenadas} to study the behavior with respect to the $y$ coordinate, we test the energy conditions in Figure \ref{FigCondicionesDeEnergia}. The vertical dashed lines located at $|y_0^+| \sim 1.17$ indicate the position in the bulk where the event horizon vanishes. In other words, the event horizon can extend over the range $y \in [-y_0^+, y_0^+]$. Outside the aforementioned interval, where the black hole horizon extends, both the energy density and the pressure components rapidly approach their asymptotic values previously given in Eq. \eqref{AsintotaCcosmologica}, implying that the spacetime outside the horizon is effectively AdS$_5$. On the other hand, close to the brane location (at $y = 0$), the energy density and pressure components are such that the Weak and Null Energy Conditions (WEC and NEC) are satisfied.

We can observe a region outside the brane and inside the black hole horizon where the bulk fluid exhibits unconventional behavior, as evidenced by the violation of the energy conditions. This feature is essential for the localization of the black hole near the brane. As pointed out in~ \cite{Nakas:2020sey}, if this condition were not satisfied, the brane could “leak” into the bulk. Therefore, this violation, confined to this specific interval, is also required to support the rotating geometry.

\section{Discussion and Summarize}

Motivated by the recent methodology presented in Ref. \cite{Nakas:2020sey}, which leads to the representation of the geometry of an analytic, exponentially localized five-dimensional braneworld black hole in a braneworld setup, we have adapted this methodology to represent the geometry of an analytic, exponentially localized rotating five-dimensional braneworld black hole. For this, we have employed the five-dimensional version of the Janis–Newman algorithm in Hopf coordinates. As a result, the resulting spacetime has been represented using this type of coordinates instead of the usual spherical coordinates. Remarkably, our formalism leads to the induced metric at the brane position corresponding to the usual 4D Kerr spacetime.

In our configuration there are two curvature singularities: the first located at $\rho = 0$, i.e. $z = r = 0$. The second is located at $\cos 2\chi = 1 \rightarrow \chi = 0$ together with $\rho = 0$. It is worth noting that this latter divergence, when evaluated at the brane location and after an algebraic analysis, leads to the usual divergence of the 4D Kerr solution at $r = 0$ and $\bar{\theta} = \pi/2$. However, since the first singularity at $\rho = 0$ is mandatory, independent of the value of the Hopf angular coordinate $\chi$, therefore, similarly to the static case \cite{Nakas:2020sey}, the black hole singularity is entirely confined to the 3-brane.

The bulk geometry is supported by a non-diagonal and highly non-linear energy–momentum tensor, as shown in Appendix \ref{EMComponentes}. The off-diagonal components of the energy–momentum tensor vanish asymptotically, while the asymptotic values of the diagonal components coincide with the value of a negative cosmological constant corresponding to a 5D Anti-de Sitter space. In this way, the energy–momentum tensor can be interpreted as a matter source that transitions in the $\rho$ coordinate from a non-diagonal and anisotropic structure to a vacuum distribution characterized by the 5D cosmological constant. Thus, the ambient spacetime transitions from a 5D rotating black hole to an AdS space. Therefore, the asymptote can be interpreted as a geometric boundary, such that the geometry can be thought of as an analytic, exponentially localized five-dimensional brane-world rotating black hole connected to an AdS$_5$ boundary.

We have also tested the behavior of the inner and event horizons, as well as the stationary limit hypersurfaces. The inner and outer horizons of the rotating black hole take the shape of a pancake, with their longer side extended along the brane and their shorter side stretching into the bulk (the extra-dimensional volume) over an exponentially small distance. Numerically, we have observed that the inner horizon vanishes more rapidly than the event horizon. This is a nontrivial result, since in recent years it has been discussed that the presence of an inner horizon may be associated with instability issues. Therefore, once the inner horizon disappears, a segment remains within the domain of the $y$ coordinate, where only an event horizon exists (without an inner one).

We have also determined how the stationary limit hypersurfaces extend along the bulk. We have described the characteristics of the horizons and stationary limit hypersurfaces in the spacetime at points \ref{Uno}, \ref{Dos}, \ref{Tres},\ref{Cuatro}, \ref{Cinco} and \ref{Seis} of Section \ref{SeccionHorizontes}.

Given that the energy-momentum tensor is non-diagonal and that the presence of the warp factor leads to highly non-linear components (see Appendix \ref{EMComponentes}), we have used a one-form from the dual basis such that a diagonal energy-momentum tensor is obtained for a specific choice of one of the angular coordinates. 

We have numerically tested the behavior of the energy conditions. Outside the interval where the black hole horizon extends, both the energy density and the pressure components rapidly approach their asymptotic values previously given in Eq.~\eqref{AsintotaCcosmologica}, implying that the spacetime outside the horizon is effectively AdS$_5$. On the other hand, close to the brane location ($y = 0$), the energy density and pressure components satisfy the Weak and Null Energy Conditions (WEC and NEC).

We can identify a region outside the brane but within the interval covered by the black hole horizon, where the bulk fluid exhibits unconventional behavior, as evidenced by the violation of the energy conditions. This feature is essential for the localization of the black hole near the brane. As pointed out in~\cite{Nakas:2020sey}, if this condition were not satisfied, the brane could “leak” into the bulk. Therefore, this violation, confined to this specific interval, is also required to support the rotating geometry.

\appendix

\section{A brief overview of the Randall-Sundrum model.} \label{ResenaRS}

The original line element for the Randall Sundrum model \cite{Randall:1999ee,Randall:1999vf} is given by
\begin{equation}
    ds^2 = e^{-2k|y|} \left( -dt^2 + d\vec{x}^2 \right)  + dy^2
\end{equation}
where $(t, \vec{x})$ represent the coordinates of the embedded surface and $y$ represents the extra coordinate. The constant $k$ is related to the inverse of the AdS$_5$ radius. This spacetime is globally AdS and has $\mathcal{Z}_2$ symmetry, such that in the one-brane model, there is a brane localized at $y =0$.

The previous line element can be written in conformally-flat coordinates, introducing the new coordinate $z$ via the relation
\begin{equation} \label{CambioCoordenadas}
    z = \text{sgn}(y) \frac{e^{k|y|} - 1}{k}
\end{equation}
such that  the resulting line element corresponds to that of equation \eqref{ElementodeLineaZ}.

It is important to mention that under this change of coordinates, the spacetime retains the same characteristics: it corresponds to a globally AdS space and has a brane embedded at $y \to z = 0$. This coordinate change has been used to study the embedding of a $4D$ black hole, which generates an additional singularity at $y \to z \to \infty$ \cite{Chamblin:1999by}, see equation \eqref{ElementodeLineaZ}.

\section{Components of the energy-momentum tensor} \label{EMComponentes}

\begin{align}
&T_{tt}^{(B)}=  -\frac{1}{16 \rho \left(b^2 - b^2 \cos(2 \chi) + 2 \rho^2\right)^3 \left(1 + k \cos(\theta) \rho \sin(\chi)\right)^2} \cdot \nonumber \\
&\bigg (  -16 b^4 M + 240 b^6 k^2 \rho + 32 b^2 M^2 \rho - 24 b^6 k^2 \cos(6 \chi) \rho - 224 b^2 M \rho^2 - 332 b^4 k^2 M \rho^2 \nonumber \\
&- 44 b^4 k^2 M \cos(2 \theta) \rho^2 + b^4 k^2 M \cos(2 \theta - 6 \chi) \rho^2 - 10 b^4 k^2 M \cos(2 \theta - 4 \chi) \rho^2 \nonumber \\
&+ 31 b^4 k^2 M \cos(2 \theta - 2 \chi) \rho^2 + 2 b^4 k^2 M \cos(6 \chi) \rho^2 + 31 b^4 k^2 M \cos(2 (\theta + \chi)) \rho^2 \nonumber \\
&- 10 b^4 k^2 M \cos(2 \theta + 4 \chi) \rho^2 + b^4 k^2 M \cos(2 \theta + 6 \chi) \rho^2 + 864 b^4 k^2 \rho^3 \nonumber \\
&+ 192 M^2 \rho^3 + 48 b^2 k^2 M^2 \rho^3 + 48 b^2 k^2 M^2 \cos(2 \theta) \rho^3 + 8 b^2 k^2 M^2 \cos(2 \theta - 4 \chi) \rho^3 \nonumber \\
&- 32 b^2 k^2 M^2 \cos(2 \theta - 2 \chi) \rho^3 - 32 b^2 k^2 M^2 \cos(2 (\theta + \chi)) \rho^3 + 8 b^2 k^2 M^2 \cos(2 \theta + 4 \chi) \rho^3 \nonumber \\
&- 192 M \rho^4 - 792 b^2 k^2 M \rho^4 - 24 b^2 k^2 M \cos(2 \theta) \rho^4 - 20 b^2 k^2 M \cos(2 \theta - 4 \chi) \rho^4 \nonumber \\
&+ 32 b^2 k^2 M \cos(2 \theta - 2 \chi) \rho^4 + 32 b^2 k^2 M \cos(2 (\theta + \chi)) \rho^4 - 20 b^2 k^2 M \cos(2 \theta + 4 \chi) \rho^4 \nonumber \\
&+ 1152 b^2 k^2 \rho^5 - 768 k^2 M \rho^6 + 768 k^2 \rho^7 + 4 b^2 \cos(4 \chi) \left(36 b^4 k^2 \rho + 2 k^2 M (2 M - 5 \rho) \rho^3 \right. \nonumber \\
&+ \left. b^2 (-4 M - 29 k^2 M \rho^2 + 72 k^2 \rho^3)\right) - 2 b^2 \cos(2 \chi) \left(180 b^4 k^2 \rho + b^2 (576 k^2 \rho^3 - M (16 + 223 k^2 \rho^2)) \right. \nonumber \\
&+ \left. 16 \rho \left(36 k^2 \rho^4 + M^2 (1 + 2 k^2 \rho^2) - M (\rho + 26 k^2 \rho^3)\right)\right) \nonumber \\
&- 16 b^4 k M \rho \sin(\theta - 5 \chi) + 72 b^4 k M \rho \sin(\theta - 3 \chi) - 56 b^2 k M^2 \rho^2 \sin(\theta - 3 \chi) \nonumber \\
&+ 104 b^2 k M \rho^3 \sin(\theta - 3 \chi) - 136 b^4 k M \rho \sin(\theta - \chi) + 168 b^2 k M^2 \rho^2 \sin(\theta - \chi) \nonumber \\
&- 408 b^2 k M \rho^3 \sin(\theta - \chi) + 288 k M^2 \rho^4 \sin(\theta - \chi) - 288 k M \rho^5 \sin(\theta - \chi) \nonumber \\
&+ 136 b^4 k M \rho \sin(\theta + \chi) - 168 b^2 k M^2 \rho^2 \sin(\theta + \chi) + 408 b^2 k M \rho^3 \sin(\theta + \chi) \nonumber \\
&- 288 k M^2 \rho^4 \sin(\theta + \chi) + 288 k M \rho^5 \sin(\theta + \chi) - 72 b^4 k M \rho \sin(\theta + 3 \chi) \nonumber \\
&+ 56 b^2 k M^2 \rho^2 \sin(\theta + 3 \chi) - 104 b^2 k M \rho^3 \sin(\theta + 3 \chi) + 16 b^4 k M \rho \sin(\theta + 5 \chi) \bigg )
\end{align}

\begin{align}
   & T_{\rho \rho}^{(B)} = -\frac{1}{8 \rho \left(b^2 - b^2 \cos(2 \chi) + 2 \rho^2\right) \left(b^2 + \rho \left(-M + \rho\right)\right) \left(1 + k \cos(\theta) \rho \sin(\chi)\right)^2} \cdot \nonumber \\
   & \bigg ( 4 b^2 M - 36 b^4 k^2 \rho + 24 M \rho^2 + 6 b^2 k^2 M \rho^2 + 6 b^2 k^2 M \cos(2 \theta) \rho^2 \nonumber \\
& + b^2 k^2 M \cos(2 \theta - 4 \chi) \rho^2 - 4 b^2 k^2 M \cos(2 \theta - 2 \chi) \rho^2 - 4 b^2 k^2 M \cos(2 (\theta + \chi)) \rho^2 \nonumber \\
& + b^2 k^2 M \cos(2 \theta + 4 \chi) \rho^2 - 96 b^2 k^2 \rho^3 - 96 k^2 \rho^5 \nonumber \\
& - 2 b^2 k^2 \cos(4 \chi) \rho (6 b^2 - M \rho) + 4 b^2 \cos(2 \chi) (12 k^2 \rho (b^2 + 2 \rho^2) - M (1 + 2 k^2 \rho^2)) \nonumber \\
& - 7 b^2 k M \rho \sin(\theta - 3 \chi) + 21 b^2 k M \rho \sin(\theta - \chi) + 36 k M \rho^3 \sin(\theta - \chi) \nonumber \\
& - 21 b^2 k M \rho \sin(\theta + \chi) - 36 k M \rho^3 \sin(\theta + \chi) + 7 b^2 k M \rho \sin(\theta + 3 \chi) \bigg )
\end{align}

\begin{align}
  & T_{\chi \chi}^{(B)} =  \frac{1}{8 \rho \left( b^2 - b^2 \cos(2 \chi) + 2 \rho^2 \right) \left( 1 + k \cos(\theta) \rho \sin(\chi) \right)^2} \cdot \nonumber \\
  & \bigg ( -8 b^2 M + 36 b^4 k^2 \rho - 3 b^2 k^2 M \rho^2 - 
3 b^2 k^2 M \cos(2 \theta) \rho^2 - \frac{1}{2} b^2 k^2 M \cos(2 \theta - 4 \chi) \rho^2 + \nonumber \\
& 2 b^2 k^2 M \cos(2 \theta - 2 \chi) \rho^2 + 
2 b^2 k^2 M \cos(2 (\theta + \chi)) \rho^2 - 
\frac{1}{2} b^2 k^2 M \cos(2 \theta + 4 \chi) \rho^2 + \nonumber \\
& 96 b^2 k^2 \rho^3 - 12 k^2 M \rho^4 - 
12 k^2 M \cos(2 \theta) \rho^4 + 
6 k^2 M \cos(2 \theta - 2 \chi) \rho^4 + \nonumber \\
& 6 k^2 M \cos(2 (\theta + \chi)) \rho^4 + 96 k^2 \rho^5 + 
b^2 k^2 \cos(4 \chi) \rho (12 b^2 - M \rho) + \nonumber \\
& \cos(2 \chi) (-48 b^4 k^2 \rho + 12 k^2 M \rho^4 + 
4 b^2 (2 M + k^2 M \rho^2 - 24 k^2 \rho^3)) + \nonumber \\
& 8 b^2 k M \rho \sin(\theta - 3 \chi) - 
24 b^2 k M \rho \sin(\theta - \chi) - 
24 k M \rho^3 \sin(\theta - \chi) + \nonumber \\
& 24 b^2 k M \rho \sin(\theta + \chi) + 
24 k M \rho^3 \sin(\theta + \chi) - 
8 b^2 k M \rho \sin(\theta + 3 \chi) \bigg )
\end{align}

\begin{align}
   & T_{\theta \theta}^{(B)} = \frac{1}{4 \left( b^2 - b^2 \cos(2 \chi) + 2 \rho^2 \right) \left( 1 + k \cos(\theta) \rho \sin(\chi) \right)^2} \cdot \nonumber \\
   & \bigg (3 k \rho^2 \sin(\chi)^2 \left( 8 b^2 k - 2 k M \rho - 2 k M \cos(2 \theta) \rho + k M \cos(2 \theta - 2 \chi) \rho \right. \nonumber \\
& \left. + k M \cos(2 (\theta + \chi)) \rho + 16 k \rho^2 + \cos(2 \chi) \left( -8 b^2 k + 2 k M \rho \right) \right. \nonumber \\
& \left. - 4 M \sin(\theta - \chi) + 4 M \sin(\theta + \chi) \right) \bigg )
\end{align}

\begin{align}
 & T_{\phi \phi}^{(B)} =  \frac{\cos^2 (\chi) }{8 \rho \left( b^2 - b^2 \cos(2 \chi) + 2 \rho^2 \right)^3 \left( 1 + k \cos(\theta) \rho \sin(\chi) \right)^2}  \cdot \nonumber \\
 & \bigg ( -28 b^6 M + 120 b^8 k^2 \rho - 4 b^4 M^2 \rho - 12 b^8 k^2 \cos(6 \chi) \rho - 12 b^4 M \rho^2 + 16 b^6 k^2 M \rho^2 - 8 b^6 k^2 M \cos(2 \theta) \rho^2 \nonumber \\
&+ b^6 k^2 M \cos(2 \theta - 6 \chi) \rho^2 - 4 b^6 k^2 M \cos(2 \theta - 4 \chi) \rho^2 + 7 b^6 k^2 M \cos(2 \theta - 2 \chi) \rho^2 + 14 b^6 k^2 M \cos(6 \chi) \rho^2 \nonumber \\
&+ 7 b^6 k^2 M \cos(2 (\theta + \chi)) \rho^2 - 4 b^6 k^2 M \cos(2 \theta + 4 \chi) \rho^2 + b^6 k^2 M \cos(2 \theta + 6 \chi) \rho^2 + 552 b^6 k^2 \rho^3 - 48 b^2 M^2 \rho^3 -\nonumber \\
& 4 b^4 k^2 M^2 \rho^3 - 4 b^4 k^2 M^2 \cos(2 \theta) \rho^3 - b^4 k^2 M^2 \cos(2 \theta - 6 \chi) \rho^3 + 2 b^4 k^2 M^2 \cos(2 \theta - 4 \chi) \rho^3 + b^4 k^2 M^2 \cos(2 \theta - 2 \chi)  \nonumber \\
&\cdot \rho^3 - 12 b^6 k^2 \cos(6 \chi) \rho^3 - 2 b^4 k^2 M^2 \cos(6 \chi) \rho^3 + b^4 k^2 M^2 \cos(2 (\theta + \chi)) \rho^3 + 2 b^4 k^2 M^2 \cos(2 \theta + 4 \chi) \rho^3 \nonumber \\
&- b^4 k^2 M^2 \cos(2 \theta + 6 \chi) \rho^3 + 16 b^2 M \rho^4 + 28 b^4 k^2 M \rho^4 - 68 b^4 k^2 M \cos(2 \theta) \rho^4 + b^4 k^2 M \cos(2 \theta - 6 \chi) \rho^4 \nonumber \\
&- 6 b^4 k^2 M \cos(2 \theta - 4 \chi) \rho^4 + 39 b^4 k^2 M \cos(2 \theta - 2 \chi) \rho^4 + 2 b^4 k^2 u \cos(6 \chi) \rho^4 + 39 b^4 k^2 M \cos(2 (\theta + \chi)) \rho^4 \nonumber \\
&- 6 b^4 k^2 M \cos(2 \theta + 4 \chi) \rho^4 + b^4 k^2 M \cos(2 \theta + 6 \chi) \rho^4 + 1008 b^4 k^2 \rho^5 + 84 b^2 k^2 M \rho^6 - 108 b^2 k^2 M \cos(2 \theta) \rho^6 \nonumber \\
&- 2 b^2 k^2 M \cos(2 \theta - 4 \chi) \rho^6 + 56 b^2 k^2 M \cos(2 \theta - 2 \chi) \rho^6 + 56 b^2 k^2 M \cos(2 (\theta + \chi)) \rho^6 - 2 b^2 k^2 M \cos(2 \theta + 4 \chi) \rho^6 \nonumber \\
&+ 960 b^2 k^2 \rho^7 - 48 k^2 M \rho^8 - 48 k^2 M \cos(2 \theta) \rho^8 + 24 k^2 M \cos(2 \theta - 2 \chi) \rho^8 + 24 k^2 M \cos(2 (\theta + \chi)) \rho^8 \nonumber \\
&+ 384 k^2 \rho^9 + 2 \cos(2 \chi) \left( -90 b^8 k^2 \rho + 24 k^2 M \rho^8 \right. \left. + b^4 \rho^2 \left( 56 M + k^2 M^2 \rho + 39 k^2 M \rho^2 - 576 k^2 \rho^3 \right) \right. \nonumber \\
&\left. + b^6 \left( 16 M + k^2 M \rho^2 - 378 k^2 \rho^3 \right) - 8 b^2 \rho^3 \left( 3 u^2 - 5 M \rho \right. \right. \left. \left. - 19 k^2 M \rho^3 + 36 k^2 \rho^4 \right) \right) \nonumber \\
&+ 4 b^2 \cos(4 \chi) \left( 18 b^6 k^2 \rho - k^2 M \rho^6 - b^4 \left( M + 8 k^2 M \rho^2 - 54 k^2 \rho^3 \right) \right. \left. + b^2 \rho \left( 36 k^2 \rho^4 + M^2 (1 + k^2 \rho^2) - M (\rho +  \right. \right. \nonumber \\
&\left. \left. 27 k^2 \rho^3) \right) \right)- 7 b^6 k M \rho \sin(\theta - 5 \chi) + 7 b^4 k M^2 \rho^2 \sin(\theta - 5 \chi) - 7 b^4 k M \rho^3 \sin(\theta - 5 \chi) + 39 b^6 k M \rho \sin(\theta - 3 \chi) \nonumber \\
&- 7 b^4 k M^2 \rho^2 \sin(\theta - 3 \chi) + 107 b^4 k M \rho^3 \sin(\theta - 3 \chi) - 36 b^2 k M^2 \rho^4 \sin(\theta - 3 \chi) + 68 b^2 k M \rho^5 \sin(\theta - 3 \chi) \nonumber \\
&- 82 b^6 k M \rho \sin(\theta - \chi) - 14 b^4 k M^2 \rho^2 \sin(\theta - \chi) - 238 b^4 k M \rho^3 \sin(\theta - \chi) - 36 b^2 k M^2 \rho^4 \sin(\theta - \chi) \nonumber \\
&- 252 b^2 k M \rho^5 \sin(\theta - \chi) - 96 k M \rho^7 \sin(\theta - \chi) + 82 b^6 k M \rho \sin(\theta + \chi) + 14 b^4 k M^2 \rho^2 \sin(\theta + \chi) \nonumber \\
&+ 238 b^4 k M \rho^3 \sin(\theta + \chi) + 36 b^2 k M^2 \rho^4 \sin(\theta + \chi) + 252 b^2 k M \rho^5 \sin(\theta + \chi) + 96 k M \rho^7 \sin(\theta + \chi) \nonumber \\
&- 39 b^6 k M \rho \sin(\theta + 3 \chi) + 7 b^4 k M^2 \rho^2 \sin(\theta + 3 \chi) - 107 b^4 k M \rho^3 \sin(\theta + 3 \chi) + 36 b^2 k M^2 \rho^4 \sin(\theta + 3 \chi) \nonumber \\
&- 68 b^2 k M \rho^5 \sin(\theta + 3 \chi) + 7 b^6 k M \rho \sin(\theta + 5 \chi) - 7 b^4 k M^2 \rho^2 \sin(\theta + 5 \chi) + 7 b^4 k M \rho^3 \sin(\theta + 5 \chi) \bigg )
\end{align}

\begin{align}
& T_{t \phi}^{(B)} =  -\frac{b M \cos^2 (\chi)}{   4 \rho \left( b^2 - b^2 \cos(2 \chi) + 2 \rho^2 \right)^3 \left( 1 + k \cos(\theta) \rho \sin(\chi) \right)^2} \cdot \nonumber \\
&\bigg (-8 b^4 - 8 b^2 M \rho + 40 b^2 \rho^2 + 78 b^4 k^2 \rho^2 
b^4 k^2 \cos(2 \theta - 4 \chi) \rho^2 - 4 b^4 k^2 \cos(2 \theta - 2 \chi) \rho^2 + \nonumber \\
&26 b^4 k^2 \cos(4 \chi) \rho^2 - 4 b^4 k^2 \cos(2 (\theta + \chi)) \rho^2 + b^4 k^2 \cos(2 \theta + 4 \chi) \rho^2 - 48 M \rho^3 - 12 b^2 k^2 M \rho^3 - \nonumber \\
&2 b^2 k^2 M \cos(2 \theta - 4 \chi) \rho^3 + 8 b^2 k^2 M \cos(2 \theta - 2 \chi) \rho^3 - 4 b^2 k^2 M \cos(4 \chi) \rho^3 + 8 b^2 k^2 M \cos(2 (\theta + \chi)) \rho^3 - \nonumber \\
&2 b^2 k^2 M \cos(2 \theta + 4 \chi) \rho^3 + 48 \rho^4 + 174 b^2 k^2 \rho^4 + b^2 k^2 \cos(2 \theta - 4 \chi) \rho^4 + 8 b^2 k^2 \cos(2 \theta - 2 \chi) \rho^4 + \nonumber \\
&2 b^2 k^2 \cos(4 \chi) \rho^4 + 8 b^2 k^2 \cos(2 (\theta + \chi)) \rho^4 + b^2 k^2 \cos(2 \theta + 4 \chi) \rho^4 + 168 k^2 \rho^6 + \nonumber \\
& 12 k^2 \cos(2 \theta - 2 \chi) \rho^6 + 12 k^2 \cos(2 (\theta + \chi)) \rho^6 +  6 k^2 \cos(2 \theta) \rho^2 (b^4 - 4 \rho^4 - b^2 \rho (2 M + 3 \rho)) + \nonumber \\
& 8 \cos(2 \chi) (3 k^2 \rho^6 + b^4 (1 - 13 k^2 \rho^2) + b^2 \rho (M + \rho + 2 k^2 M \rho^2 - 22  k^2 \rho^3)) +  2 b^4 k \rho \sin(\theta - 3 \chi) + \nonumber \\ 
& 14 b^2 k M \rho^2 \sin(\theta - 3 \chi) +  2 b^2 k \rho^3 \sin(\theta - 3 \chi) - 6 b^4 k \rho \sin(\theta - \chi) -  42 b^2 k M \rho^2 \sin(\theta - \chi) + \nonumber \\
& 18 b^2 k \rho^3 \sin(\theta - \chi) -  72 k M \rho^4 \sin(\theta - \chi) + 24 k \rho^5 \sin(\theta - \chi) +  6 b^4 k \rho \sin(\theta + \chi) + 42 b^2 k M \rho^2 \sin(\theta + \chi) - \nonumber \\
& 18 b^2 k \rho^3 \sin(\theta + \chi) + 72 k M \rho^4 \sin(\theta + \chi) -  24 k \rho^5 \sin(\theta + \chi) - 2 b^4 k \rho \sin(\theta + 3 \chi) - \nonumber \\
& 14 b^2 k M \rho^2 \sin(\theta + 3 \chi) - 2 b^2 k \rho^3 \sin(\theta + 3 \chi) \bigg )
\end{align}

\section*{Acknowledgements}
 Milko Estrada is funded by ANID , FONDECYT de Iniciaci\'on en Investigación 2023, Folio 11230247. 

\bibliography{mybib.bib}

\end{document}